# Cyclic stress-dilatancy relations and associated flow for soils based on hypothesis of complementarity of stress-dilatancy conjugates

Anteneh Biru Tsegaye

Norwegian Geotechnical Institute, anteneh.biru.tsegaye@ngi.no

## Abstract

*Applicability of associated plasticity for particulate materials such as soils does not yield satisfactory results when Coulomb's theory of shear strength of soils is assumed, and the yield function derived accordingly is used to define both the stress state and the direction of plastic flow. The limitation mainly stems from the fact that Coulomb's theory (and its derivatives) is a simplification that intentionally ignores deformation characteristics that manifest from the particulate nature of such materials. It is thus customary to apply a branch of plasticity called non-associated plasticity for soils and similar materials. In the non-associated plasticity framework, yield functions and plastic potential functions are different. The former defines the mobilization of the stress state while the later defines the direction of plastic flow. For soils, stress-dilatancy theories have become central in the formulation of non-associated flow rules. In this paper, cyclic stress-dilatancy relations are derived based on complementarity hypothesis of stress-dilatancy conjugates. Both loading and unloading are explicitly considered. Then, yield functions are derived based on the resulting stress-dilatancy relations. In so doing, the resulting yield functions are rendered with a quality to be used for the modelling of deformation behavior of soils subjected to monotonic and cyclic loading conditions. The newly derived yield functions are called Associated Cyclic Stress Dilatancy yield functions for which the abbreviation ACStD is used. The theoretical framework is established first for special cases of deformation modes- plane strain and axisymmetric. The framework is generalized for considering Lode angle dependency of the yield function and extending the Matusoka-Nakai criterion.*

## 1. Introduction

The theory of plasticity is successfully applied for describing the deformation behavior of continuous materials such as metals. It has also been applied to discontinuous assemblies of particles such as sands and clays but some of the axioms needed to be relaxed to accommodate properties that emerge from the discontinuous nature of such materials.

Elastoplasticity builds on additive decomposition of strain rates into elastic and plastic. The decomposition of the strain rate into elastic and plastic requires that each component be described with certain assumptions. The elastic portion of the strain rate is assumed to be uniquely determined by the corresponding stress increments, where the elastic moduli are assumed known. For metals, a positive definite elastic stiffness tensor with constant moduli is often assumed. For soils, rocks and similar materials, the elastic moduli are assumed to depend on several factors such as the void ratio, effective confining pressure and history of loading [1]. In addition, the definition of elasticity for discontinuous materials such as soils, is somewhat arbitrary. On the other hand, the plastic strain rate is assumed to depend on the current stress state in some way (not on the stress increment). This property was originally postulated by Saint-Venant [2]. Later, von Mises proposed a potential function of stress whose gradient with respect to stresses is assumed to give the direction of plastic flow. Assumption of coaxiality of principal stresses and principal plastic strain rates was then introduced. Drucker [3] presented his postulate of material stability in which the plastic potential function has to be identical to the yield function, which a stress state obeys and that the yield function has to be convex. This branch of elastoplasticity where the yield function serves as a plastic potential function as well is called associated plasticity. Even though associated plasticity is a well-established theoretical framework with several elegant mathematical theorems and its success in its application for metals, with yield functions that are based on Coulomb's shear strength theory, the application of associated plasticity to soils, rocks and concrete give unrealistic plastic volumetric strain during plastic deformation [4]. This necessitated establishing a plastic potential function that is different from the yield function for realistic prediction of shear induced plastic volumetric strains. This second type of plasticity framework is called non-associated plasticity. Technically, associated plasticity will then be a special class of non-associated plasticity. The limitation of associated plasticity to reproduce the observed deformation behavior of soils and similar materials is often identified as a limitation of the framework itself. This is not entirely true. For the most part, it is the limitation of Coulomb's formula for the description of shear strength of soils- which is also carried over to several of its extended forms.

Coulomb [5] described the shear strength of soils in terms of their friction angle and cohesion as[1]

$$\tau = \overset{1}{\sigma_n \tan\varphi + c} = \overset{2}{(\sigma_n + a)\tan\varphi} \tag{1}$$

where $\varphi$ is the friction angle or angle of internal friction, $c$ is cohesion and $a = c \cot \varphi$ is attraction [6], Figure 1. Coulomb's shear strength theory held a central place in traditional earth-pressure theories, but it ignores the fact that soils contain grains [7]-which renders them with certain peculiar properties when sheared. Reynolds [8] recognized the limitation of Coulomb's shear strength theory after his discovery of the property of dilatancy-which manifests from the particulate nature of sands. Reynolds then envisaged that the consideration of the property of dilatancy will place earth-pressure theories on a true foundation.

The first attempts to apply the principles of dilatancy to earth pressure problems appear in Jenkin [9]. Shortly after, Casagrande [10] discussed the dilatancy behavior of sands in line with his critical state void ratio (porosity) concept. One of the earliest attempts to describe dilatancy in terms of the energy dissipation was due to Taylor [11]. The work of Taylor has been a basis of various dissipation equations and stress-dilatancy formalisms. The other interesting theoretical framework for describing the relationship between stress ratio and dilatancy ratio is due to Rowe [12]. Taylor's and Rowe's theories have been modified, extended and reinterpreted in the literature [13]. There exist a vast body of literature on the stress-dilatancy behavior of geomaterials, experimental as well as theoretical. As Jefferies and Been [14] stated, the stress-dilatancy framework is "a basic and excellent framework for understanding soil" mechanical behavior" which the author of this paper fully agrees to.

The objective of this paper is establishing yield functions in the framework of associated plasticity based on the Hypothesis of Complementarity of Stress-dilatancy conjugates (HCSDC) put forward further down in the paper. Accordingly, first plastic dissipation is discussed in its generality and the role of stress-dilatancy relations in the plastic dissipation is pointed out. Then, a yield function which defines both the mobilization of friction and plastic flow defined by stress-dilatancy relations are derived. The theoretical framework is presented first for axisymmetric and plane strain conditions and then extended for the general stress-strain conditions.

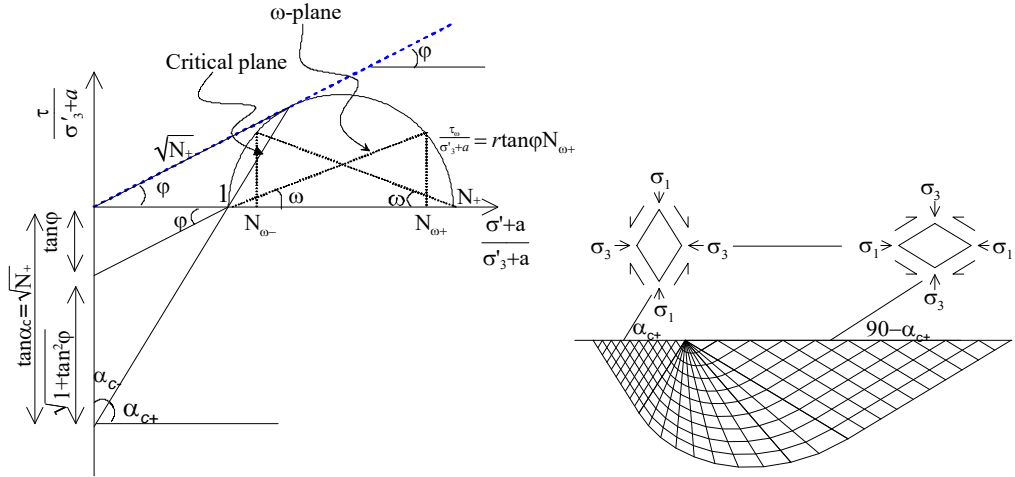

**Figure 1**: (a) Normalized Mohr's stress circle, dimensionless quantities and Coulomb's criterion [15], (b): Illustration in classical geometry of bearing capacity mechanism for a vertically loaded foundation and critical element in the active Rankine and passive Rankine zones. $\sigma_1$ and $\sigma_3$ are, respectively, the major and the minor principal stresses [1].

## 2. Plastic dissipation, non-coaxiality and stress-dilatancy conjugates

For an isothermal condition, the average energy variation in a deforming body may be written as

$$\dot{F} + \mathcal{D} - W = 0, \quad \mathcal{D} \geq 0 \tag{2}$$

where $\dot{F}$ is the rate of Helmholtz free energy, $\mathcal{D}$ is the rate of dissipation and $W$ is the rate of work.

Considering the additive decomposition of the strain rate into elastic and plastic, the rate of work of a continuum body may be written as

$$W = \sigma_{ij}\dot{\varepsilon}_{ij} = \sigma_{ij}\dot{\varepsilon}^e_{ij} + \sigma_{ij}\dot{\varepsilon}^p_{ij} = p\dot{\varepsilon}^e_v + c^e_\Delta q\dot{\varepsilon}^e_q + p\dot{\varepsilon}^p_v + c^p_\Delta sq\dot{\varepsilon}^p_q \tag{3}$$

where

- $\sigma_{ij}$ is Cauchy's tress tensor, $p = \frac{1}{3}\sigma_{ij}\delta_{ij}$ is the isotropic stress, also called mean effective stress, effective confining stress or effective confining pressure, $\delta_{ij}$ is Kronecker's delta with a property $\delta_{ij} = 1, i = j$ and $\delta_{ij} = 0, i \neq j$ and Einstein's summation rule over repeated indices applies.
- $s_{ij} = \sigma_{ij} - \sigma_{ij}\delta_{ij}/3$ is the deviatoric stress tensor, it is a traceless tensor that contains all the shear components. The magnitude of the deviatoric stress is given by $||s|| = \sqrt{s_{ij}s_{ij}} = q\sqrt{2/3}$, and $q$ is called deviatoric stress.
- $\dot{\varepsilon}_v = \dot{\varepsilon}_{ij}\delta_{ij}$ is the volumetric strain rate.
- $\dot{e}_{ij}$ is distortional or deviatoric strain rate tensor which is the deviation from mean isotropic straining. It is thus obtained by subtracting the mean normal strain rate from the total strain rate tensor as $\dot{e}_{ij} = \dot{\varepsilon}_{ij} - \dot{\varepsilon}_v\delta_{ij}/3$ and its magnitude is given by $||\dot{e}|| = \sqrt{\dot{e}_{ij}\dot{e}_{ij}} = \sqrt{3/2}\,\dot{\varepsilon}_q$, where $\dot{\varepsilon}_q$ is called the deviatoric strain rate.
- $c_\Delta$ is the degree of coaxiality between the respective stress and strain increments ($c_\Delta = 1$ when they are coaxial.)
- $s = 1$ for shear loading (loading away from isotropic condition) and $s = -1$ for shear unloading (loading towards isotropic condition)
- The superscripts $e$ and $p$ respectively indicate elastic and plastic.

Elastic strain increments are assumed to be coaxial with stress increments, *i.e.*,

$$c_\Delta^e = \sqrt{3/2}\,\frac{s_{ij}\dot{s}_{ij}}{q\,\|\dot{s}_{ij}\|} = \sqrt{2/3}\,\frac{\dot{q}}{\|\dot{s}_{ij}\|}, \text{ for } \|\dot{s}_{ij}\| > 0\,. \quad (4)$$

Let us define, according to Ishihara and Gutierrez [16],

$$s_{ij} = q\tilde{m}_{ij} \text{ and } \dot{e}_{ij}^p = \dot{\varepsilon}_q^p \tilde{n}_{ij}, \quad (5)$$

such that

$$c_\Delta^p = \|\tilde{m}_{ij}\tilde{n}_{ij}\|,\ s = \text{sgn}(\tilde{m}_{ij}\tilde{n}_{ij}), \quad (6)$$

where $\tilde{m}_{ij}$ and $\tilde{n}_{ij}$ are respectively given by

$$\tilde{m}_{ij} = \frac{2}{3}\left[T_{i1}^\sigma T_{j1}^\sigma \sin\left(\theta_\sigma + 4\pi/3\right) + T_{i2}^\sigma T_{j2}^\sigma \sin\left(\theta_\sigma + 2\pi/3\right) + T_{i2}^\sigma T_{j2}^\sigma \sin\left(\theta_\sigma\right)\right], \text{ and} \quad (7)$$

$$\tilde{n}_{ij} = T_{i1}^{\dot{\varepsilon}^p} T_{j1}^{\dot{\varepsilon}^p} \sin\left(\theta_{\dot{\varepsilon}}^p + 4\pi/3\right) + T_{i2}^{\dot{\varepsilon}^p} T_{j2}^{\dot{\varepsilon}^p} \sin\left(\theta_{\dot{\varepsilon}}^p + 2\pi/3\right) + T_{i2}^{\dot{\varepsilon}^p} T_{j2}^{\dot{\varepsilon}^p} \sin\left(\theta_{\dot{\varepsilon}}^p\right). \quad (8)$$

$T_{ik}^\sigma$ and $T_{ik}^{\dot{\varepsilon}}$ are matrices that transform the stress tensor and the plastic strain rate tensors into their respective principals, $\theta_\sigma$ is the Lode angle of the stress tensor; $\theta_{\dot{\varepsilon}}^p$ is the Lode angle of the plastic strain rate tensor. If $T_{ik}^\sigma \neq T_{ik}^{\dot{\varepsilon}}$, then the principal stresses and principal strain rates are said to be non-coaxial and the condition is referred to as non-coaxiality.

Coaxiality between principal stresses and principal plastic strain rates was first postulated by Saint-Venant [2]. The assumption of coaxiality between stresses and plastic strain increments has however been contested. It has been pointed out by Hill [17] that for anisotropic material, generally principal stresses and principal (plastic) strain increments are non-coaxial except for the special case where the principal stress axes coincide with the axes of anisotropy. In fact, in granular materials, non-coaxiality between principal stresses and principal (plastic) strain rates has been observed through various techniques. For example, Drescher and de Josselin de Jong [18], studied the deformation behaviour of a photo elastic disc assembly to verify the double-sliding free-rotating model[1] [19] and they were able to calculate the degree of non-coaxiality between the axes of principal stresses and strain rates. Roscoe *et al.* [20], using simple shear tests, observed that principal stresses and principal plastic strain rates can be non-coaxial. Using the directional shear cell apparatus (DSC) [21], Arthur *et al.* [22] investigated the stress-strain behaviour of the Leighton-Buzzard Sand samples due to change of stress path direction and found that principal stresses and principal strain rates are generally non-coaxial. Gutierrez *et al.* [23] employed the hollow cylinder apparatus to investigate deformation behaviour of dense air-pulviated Toyoura sand subjected to proportional stress path, pure principal stress rotation and loading with increasing deviatoric stress combined with principal stress rotation. Their test results show that loading conditions that involve principal stress rotations are in general non-coaxial. The hollow cylinder has been popularly applied to the investigation of stress-strain behaviour of soils under loading conditions that involve principal stress rotation *e.g.*, Cai, *et al.* [24]. Non-coaxiality has also been observed in discrete element model (DEM) set ups [22, 25-27].

In this paper, we will not consider non-coaxiality in more details. We will rather focus on the stress-dilatancy relationship. Let us focus on the plastic part of the energy rate in Equation (2). We assume that the total plastic dissipation is due to mobilization of friction, dilation and attraction. Suppose $\dot{F}$ is decomposed into elastic ($\dot{F}^e$) and plastic ($\dot{F}^p$) parts such that the elastic part of the free energy rate is equal to the elastic rate of work, i.e., $\dot{F}^e = \sigma_{ij}\dot{\varepsilon}^e_{ij}$, and the plastic free energy rate due to attraction, $a^2$, is considered as $\dot{F}^p := -a\dot{\varepsilon}^p_v$, considering the plastic work in Equation (3) and rearranging, the plastic dissipation, $D^p$, may be written as

$$D^p = \dot{W}^p - \dot{F}^p = (p+a)c^p_\Delta \left( \frac{\dot{\varepsilon}^p_v}{c^p_\Delta \dot{\varepsilon}^p_q} + s\frac{q}{p+a} \right), \tag{9}$$

the stress-dilatancy formalism seeks for the relationship between the stress-dilatancy conjugates in the bracket. The stress-dilatancy relationship in soils and rocks and similar materials is therefore an integral part of their energy dissipation mechanism. One of the simplest relationships is when the sum of the stress-dilatancy conjugates in the bracket is just a constant. In fact, this is the main hypothesis in classical stress-dilatancy theories. This may be phrased in a more genialized hypothesis presented below– which is here called hypothesis of complementarity of stress-dilatancy conjugates (HCSDC).

---

[1]The model was abstracted such that the system of grains is subdivided into elements that freely rotate and slide with respect to each other.

## 3. Hypothesis of complementarity of stress-dilatancy conjugates

The author proposed a theoretical framework for establishing stress-dilatancy relations considering a hypothesis that (hitherto not explicitly stated):

*for a given state, the function in the plastic rate of work that contains the stress-ratio and its conjugate dilatancy-ratio vanishes under first variation and the plastic dissipation is non-negative in both loading in shear (away from isotropy) and unloading in shear (towards isotropy*.)

This hypothesis will be referred to, in the following, as hypothesis of *complementarity of stress-dilatancy conjugates* (HCSDC). The HCSDC hypothesis yields stress-dilatancy relations depending on the choice the stress-dilatancy conjugates. The hypothesis agrees with both Taylor's work hypothesis and Rowe's minimum energy ratio hypothesis and generalizes them for the case of loading and unloading and it is also found suitable for introducing non-coaxiality between eigen directions of stresses and plastic strain rates and critical state into the stress-dilatancy theories [1, 28, 29]. The author applied the resulting stress-dilatancy relations for developing state of the art plasticity model framework named Cyclic State Dilatancy model, abbreviated CStaD [1], which aims at the modelling of deformation behavior of soils subjected to both monotonic and cyclic shear-which in its preliminary evaluation turns out to qualitatively reproduce the dilatancy behavior of soils under both monotonic and cyclic loading.

## 4. Relations between stress-dilatancy conjugates and derivation of associated plastic flow potential

We begin to employ the HCSDC in plane strain and in axisymmetric conditions and formulate a plastic potential function in the framework of associated plasticity. Then, we will apply the same approach considering the full stress and plastic strain rate tensors. The current treatment will be limited to the assumption of coaxiality between principal directions of stresses and plastic strain increments.

Note that:
- Strain rates defined in this paper refer generally to an artificial time increment and can likewise be considered infinitesimal strain increments.
- All stress quantities are effective without distinguishing them with a prime or not necessarily using the adjective "effective".

---

[2] Attraction is defined as cohesion divided by the tangent of the friction angle.

Cyclic stress-dilatancy relations and associated flow for soils...

## 4.1 Plane strain and axisymmetric

The plastic dissipation in plane strain, triaxial compression and triaxial extension conditions may be conveniently written as

$$D^p = r_1(\sigma_1 + a)\dot{\varepsilon}_1^p + r_3(\sigma_3 + a)\dot{\varepsilon}_3^p, \tag{10}$$

wherein $\sigma_i$ and $\dot{\varepsilon}_i^p$ are principal stress and plastic strain rate components respectively ($i = 1$ for major, and $i = 3$ for minor) and are assumed coaxial, $r_i$ depend on mode of shear ($r_1 = r_3 = 1$ for plane strain, $r_1 = 2r_3 = 2$ for triaxial extension and $2r_1 = r_3 = 2$ triaxial compression), and $a$ is attraction [6].

Considering Coulomb's shear strength theory in a Mohr-circle, Figure 1, the relationship between principal stress components is written as

$$\sigma_1 + a = N_\varphi(\sigma_3 + a), \tag{11}$$

where $N_\varphi$ is the stress ratio. We also assume that orthogonal plastic strain rates are related as

$$\dot{\varepsilon}_3^p = -N_\psi \dot{\varepsilon}_1^p, \tag{12}$$

where $N_\psi$ is termed here as the dilatancy ratio.

The plastic rate of work per unit volume can now be written as

$$D^p = r_1(\sigma_1 + a)\dot{\varepsilon}_1^p d_N, \tag{13}$$

where

$$d_N = 1 - m_s \frac{N_\psi}{N_\varphi}, \quad m_s = r_3/r_1. \tag{14}$$

Considering the variation according to the HCSDC [1]

$$\delta d_N = -N_\varphi \delta(m_s N_\psi) + m_s N_\psi \delta N_\varphi = 0 \tag{15}$$

yields

$$C_N m_s N_\psi = N_\varphi, \tag{16}$$

where $C_N$ is a 'constant' which may have different values for different shearing modes, fabric, sample densities. Note that, although phrased in a more advantageous form, the variation in Equation (15) is equivalent to Rowe's *minimum energy ratio* or *least work* hypothesis. Equation (16) describes a stress-dilatancy relationship, as commonly known, and $N_\sigma$ and $N_\psi$ are referred to as stress-dilatancy conjugates [1].

From Equations (10) and (16), the plastic dissipation is obtained as

$$\mathcal{D}_N^p = r_1 \left( \sigma_1 + a \right) \dot{\varepsilon}_1^p \left( \frac{C_N - 1}{C_N} \right) \geq 0. \tag{17}$$

Assuming non-negative plastic dissipation, the inequality

$$C_N = \langle -s \rangle C_N^U + \langle s \rangle C_N^L, \; s = \operatorname{sgn} \dot{\varepsilon}_1^p, \; 0 < C_N^U = 1/C_N^L \leq 1, \tag{18}$$

was proposed [1, 28] where $\langle \rangle$ is the Macaulay bracket, the superscripts $L$ and $U$ respectively indicate loading and unloading. Note that $C_N^U$ does not need to be the inverse of the $C_N^L$ but its value must be less than unity for making sure non-negative plastic dissipation during unloading. The inverse relationship is just one of the possibilities.

Considering a Mohr-Coulomb (MC) material, the mobilized stress ratio and the critical state stress ratio are defined as

$$N_\varphi^{MC} \overset{1}{=} \frac{1 + \sin \varphi_m}{1 - \sin \varphi_m} \text{ and } C_N^{L,MC} \overset{2}{=} \frac{1 + f_{sd} \sin \varphi_c}{1 - f_{sd} \sin \varphi_c}, 0 \leq f_{sd} \sin \varphi_c < 1 \tag{19}$$

in which $\varphi_m$ is the mobilized friction angle, $\varphi_c$ is the critical state friction angle, and $f_{sd}$ is an ad-hoc function introduced into Equation (19)[2] such that effect of density is taken into account.

*Shear loading*

The stress-dilatancy relationship is obtained by substituting the relations in Equations (19) into Equation (16) as

$$C_N^{L,MC} m_s N_\psi = N_\varphi^{MC} \tag{20}$$

Then, the negative of the mobilized dilatancy angle is given as [13]

$$-\sin \psi_m := \overset{1}{=} \frac{m_s N_\psi - 1}{m_s N_\psi + 1} \overset{2}{=} \frac{N_\varphi^{MC} - C_N^{L,MC}}{N_\varphi^{MC} + C_N^{L,MC}}. \tag{21}$$

After some simple rearrangement one is led to:

$$-\sin \psi_m = \frac{\sin \varphi_m - f_{sd} \sin \varphi_c}{1 - f_{sd} \sin \varphi_m \sin \varphi_c}. \tag{22}$$

Equation (22) is the enhanced form of Rowe's [12] stress-dilatancy equation proposed by Wan and Guo [30]. For $f_{sd} = 1$ and $\varphi_c = \varphi_\mu$, where $\varphi_\mu$ is interparticle friction angle, Equation (22) simplifies to the original Rowe's stress-dilatancy relationship.

Rowe [12, 31] established his stress-dilatancy relationship theory for granular materials by considering the kinematics and the stress state of a pack of orderly arranged steel rods, Figure 2. The relationship has been widely applied in constitutive models for soils either as it is or with some modifications for example in [30, 32-34] among others. It has also been re-derived from some other assumptions, for example [35, 36].

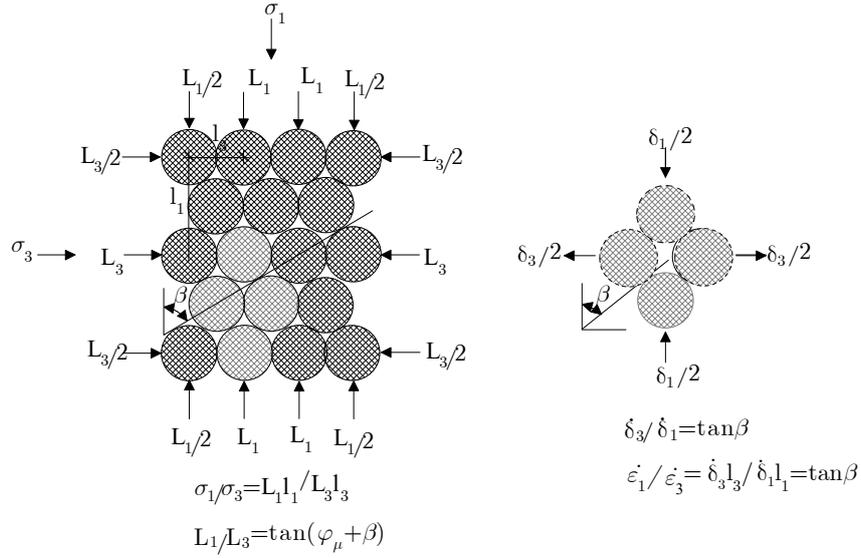

**Figure 2**: Stress and kinematics of a pack of orderly arranged cylindrical bars (after Rowe [12]).

*Shear unloading*

For the case of unloading, considering Equation (16), i.e., $C_N^U = \frac{1}{C_N^L}$ the stress-dilatancy relationship is obtained as

$$-\sin\psi_m = \frac{N_\varphi^{MC} C_N^{L,MC} - 1}{N_\varphi^{MC} C_N^{L,MC} + 1}. \tag{23}$$

Considering the definitions in Equation (19), Equation (23) simplifies to

$$-\sin\psi_m = \frac{\sin\varphi_m + f_{sd}\sin\varphi_c}{1 + f_{sd}\sin\varphi_m \sin\varphi_c}. \tag{24}$$

Note that the minus sign in Equation (22) changes into a plus sign in Equation (24). According to Equation (24), the plastic shear unloading plasticity is strictly contractive. The loading and unloading stress-dilatancy relations are thus combined as

$$-\sin\psi_m = \frac{\sin\varphi_m - sf_{sd}\sin\varphi_c}{1 + f_{sd}\sin\varphi_m \sin\varphi_c}. \tag{25}$$

where $s = 1$ during shear loading and $s = -1$ during shear unloading.

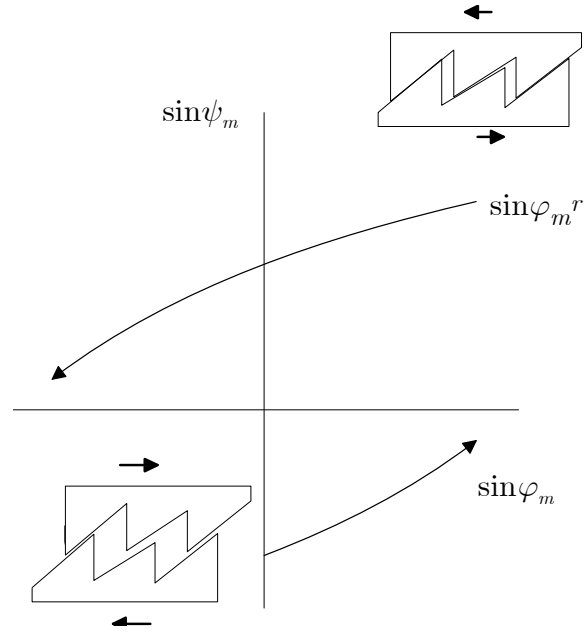

**Figure 3:** Saw blade illustration of dilatancy for loading and unloading [1].

Considering the stress-dilatancy relation in Equation (16) and further assuming associated plasticity we have

$$C_N m_s \frac{d\sigma_1}{d\sigma_3} = N_\sigma, \quad N_\psi = \frac{d\sigma_1}{d\sigma_3}. \tag{26}$$

The solution of this differential equation is:

$$\sigma_1 + a = C(\sigma_3 + a)^{\frac{1}{C_N m_s}}. \tag{27}$$

The constant of integration $C$ may be established by considering a boundary condition along the curves described by Equation (27). A boundary condition considered here is where along the curve defined by Equation (27) the stress state is isotropic, *i.e.*, $\sigma_1 = \sigma_3$. This stress we call the apparent pre-consolidation and denote it with $p_c$. At this point

$$C = (p_c + a)^{\frac{C_N m_s - 1}{C_N m_s}}. \tag{28}$$

Combining these Equations (27) and (28), we have:

$$\frac{\sigma_1 + a}{p_c + a} = \left(\frac{\sigma_3 + a}{p_c + a}\right)^{\frac{1}{C_N m_s}}. \tag{29}$$

The stress ratio, $N_\varphi$, can now be given as:

$$N_\varphi = \frac{\sigma_1 + a}{\sigma_3 + a} = \left(\frac{\sigma_3 + a}{p_c + a}\right)^{\frac{1 - C_N m_s}{C_N m_s}}. \tag{30}$$

Cyclic stress-dilatancy relations and associated flow for soils…Or, the yield function, which we call now an Associated Cyclic Stress Dilatancy yield function, abbreviated ACStD, can be defined as

$$f = \frac{\sigma_1 + a}{p_c + a} - \left(\frac{\sigma_3 + a}{p_c + a}\right)^{\frac{1}{C_N m_s}} = 0. \tag{31}$$

Considering the consistency condition, *i.e.*, $df = 0$ one arrives at:

$$\frac{d\sigma_1}{d\sigma_3} = \frac{1}{C_N m_s}\left(\frac{\sigma_3 + a}{p_c + a}\right)^{\frac{1-C_N m_s}{C_N m_s}}. \tag{32}$$

The mobilized dilatancy angle is defined as:

$$\sin\psi_m = \frac{1 - m_s N_\psi}{1 + m_s N_\psi} = \frac{(\sigma_3 + a)^{\frac{1-C_N m_s}{C_N m_s}} - C_N (p_c + a)^{\frac{1-C_N m_s}{C_N m_s}}}{(\sigma_3 + a)^{\frac{1-C_N m_s}{C_N m_s}} + C_N (p_c + a)^{\frac{1-C_N m_s}{C_N m_s}}}. \tag{33}$$

Conversely, the apparent pre-consolidation stress can also be related to the dilatancy angle as

$$p_c + a = \left(\frac{1}{C_N}\frac{1 - \sin\psi_m}{1 + \sin\psi_m}\right)^{\frac{C_N m_s}{1-C_N m_s}}(\sigma_3 + a). \tag{34}$$

Non-associated plasticity uses a yield function and a plastic potential function that are different. As we have shown here, if a yield function that is suitable for capturing both yielding and plastic flow, there is no need for making a distinction between the two. In fact, the plastic potential function defines the tangent to the actual yield function and hence the direction of the plastic strains whereas the conventional yield function mobilizes together with current stress state, Figure 4 and Figure 5. The tangent line anywhere along the ACStD yield function defined in Equation (31) is:

$$\sigma_1 = \frac{1}{C_N m_s}\left(\frac{\sigma_3 + a}{p_c + a}\right)^{\frac{1-C_N m_s}{C_N m_s}}\sigma_3 + c_a, \quad N_\psi = -\frac{1}{C_N m_s}\left(\frac{\sigma_3 + a}{p_c + a}\right)^{\frac{1-C_N m_s}{C_N m_s}} \tag{35}$$

where $c_a$ is called the apparent cohesion and is given as

$$c_a = \left(N_\sigma - N_\psi\right)\sigma_3. \tag{36}$$

As long as the apparent cohesion is derived at the point of tangency, Equation (35) may serve as both a yield function and a plastic potential function [37].

The critical state can now be investigated by considering $m_s N_\psi = 1$, which leads to

$$\sigma_{1ct} + a = \left(C_N\right)^{\frac{1}{1-C_N m_s}}(p_c + a) \text{ and } \sigma_{3ct} + a = \left(C_N\right)^{\frac{C_N m_s}{1-C_N m_s}}(p_c + a). \tag{37}$$

The stress ratio at the critical state is therefore as expected

$$\frac{\sigma_{1ct} + a}{\sigma_{3ct} + a} = C_N. \tag{38}$$

A possible form for $C_N$ proposed by the author [1] is

$$C_N = \frac{1 + f_{sd} \sin \varphi_c}{1 - f_{sd} \sin \varphi_c}. \tag{39}$$

where $f_{sd}$ is a void-ratio dependency function. The two desired properties of the void ratio dependency function are that that $f_{sd} \sin \varphi_c \in [0,1)$ and that it evolves to one when shear mobilizes towards the critical state. The phase transformation point-defined as a point where the deformation state changes from contractive to dilative can then be reached before reaching the critical state. There are several candidate functions in literature that can be used for $f_{sd}$. One possible function that can be used to define $f_{sd}$ is the Gudehus-Bauer [38, 39] state dependent function, which is given as:

$$f_{sd} = \left(\frac{e - e_d}{e_c - e_d}\right)^\alpha, \tag{40}$$

where $e_c$ and $e_d$ are respectively at the critical sate void ratio and the minimum void ratio at the current effective confining pressure. At the minimum void ratio, $C_N = 1$ is obtained. That is the stress ratio is the same as the dilatancy ratio. Introducing a non-constant $f_{sd}$ overrules the constancy we assumed. However, it seems to be the case that the stress ratio depends on the density of the sample and may be also the fabric of the sample. If $C_N$ depends on the initial void ratio, consistency demands that it must also be dependent on subsequent void ratios during shearing. The effect of fabric is a bit difficult to account to – as fabric is usually vaguely defined-and also its representation in continuum is somewhat arbitrary.

In Figure 4, the ACStD yield function is plotted for both shear loading and shear unloading for $C_N = 3$ and $p_c = 400$ kPa. The figure shows the conjugate dilatancy ratio and stress ratio as geometric properties of the ACStD yield function. The ACStD yield curve that lies above the isotropic axis are valid for stress states effective radial stress ($\sigma_r$) less than effective axial stress($\sigma_a$) while the conjugate ACStD yield curve lies bellow the isotropic axis are valid for stress states $\sigma_a < \sigma_r$. The extension of each on each conjugate yield function that extend beyond the $p_c$ are valid for unloading. For the case of loading, the stress-state contained within phase transformation lines is for contractive states where as the stress-states on the outside of the phase transformation lines is dilative. Unloading in shear is always contractive. The curves are also plotted in Figure 5 in $s - t$ space (defined in the figure) where the dilatancy angle is interpreted as a tangent to the ACStD yield function.

In Figure 6, the ACStD yield functions are plotted in $\sigma_a - \sigma_r$ space (left) and $s - t$ space (right) for both loading and unloading for a constant $C_N$ and varying values of apparent pre-consolidation stress ($p_c$). The increase in the apparent $p_c$ results in an increase in the size of the yield function. In Figure 7, the ACStD yield function is plotted for both loading and unloading conditions and for $p_c = 400$ and varying $C_N$. As can be seen, the yield curves go from relatively flat to pointed for increasing values of $C_N$. The extension of the curves beyond $p_c$ are valid only for unloading.

Let us now look at a purely isotropic compression stress state at $p = p_c$ and the consequence of the theoretical frameworks we laid so far. Let us consider the direction of plastic strain increment at $p_c$. Considering the yield function for loading alone and considering normality of the plastic strain increment vector indicates that shear induced plastic volumetric strain or the vice versa at this point. Logically, we expect purely compressive stresses to give only volumetric compression. However, with the consideration of both loading and unloading a new insight is obtained. At this point, both loading and unloading are equally legitimate and, therefore, Koiter's rule [5] may be considered such that:

$$\sin \psi_m = \sin \psi_{m;L}\Big|_{at\ p_c} + \sin \psi_{m,U}\Big|_{at\ p_c} = \frac{1-C_N}{1+C_N} + \frac{C_N-1}{1+C_N} = 0. \tag{41}$$

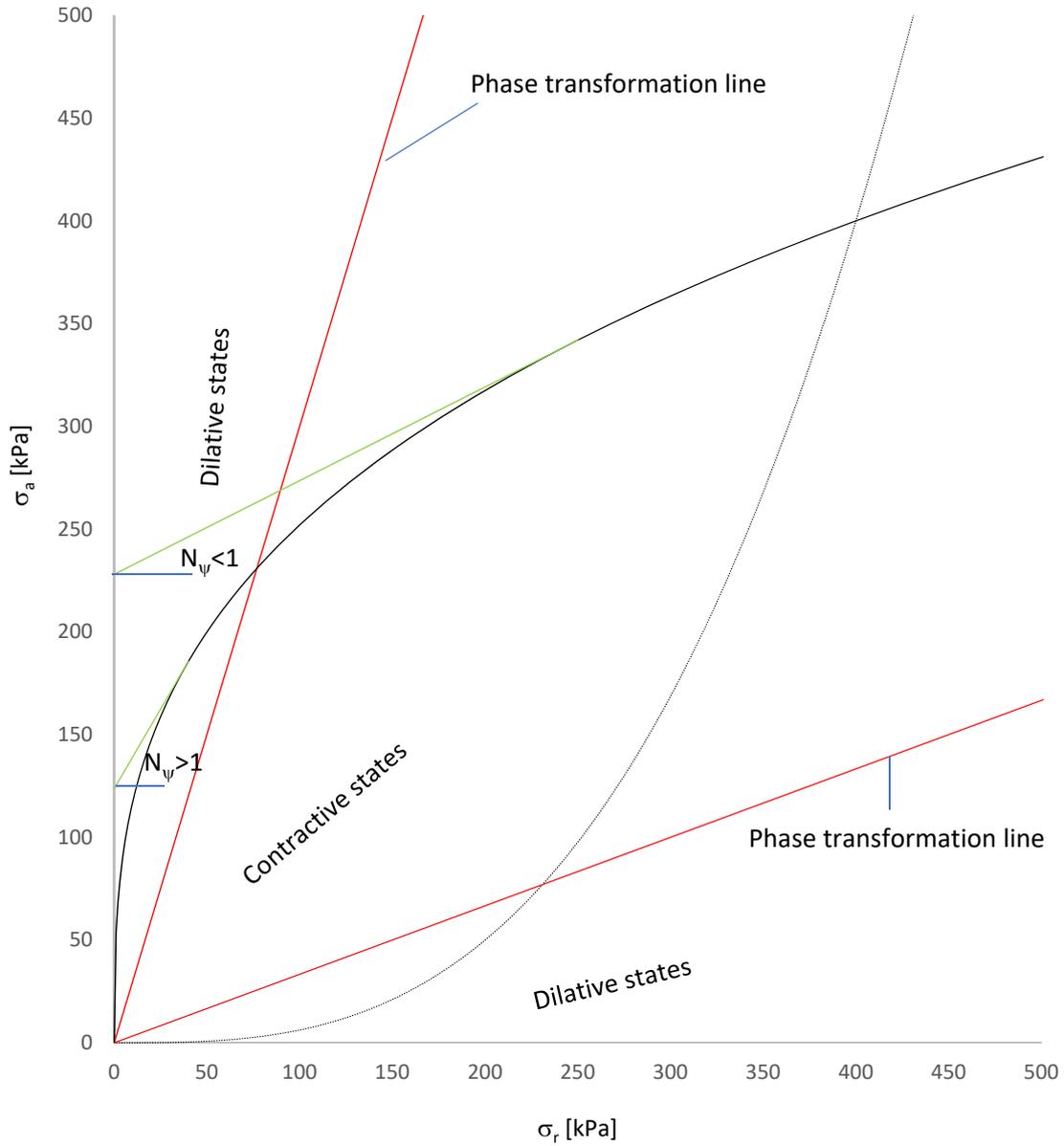

**Figure 4:** Plot of the ACStD yield curves in axial stress radial stress space for pc = 400 kPa and $C_N = 3$. The solid line is for loading and the dotted line is for unloading.

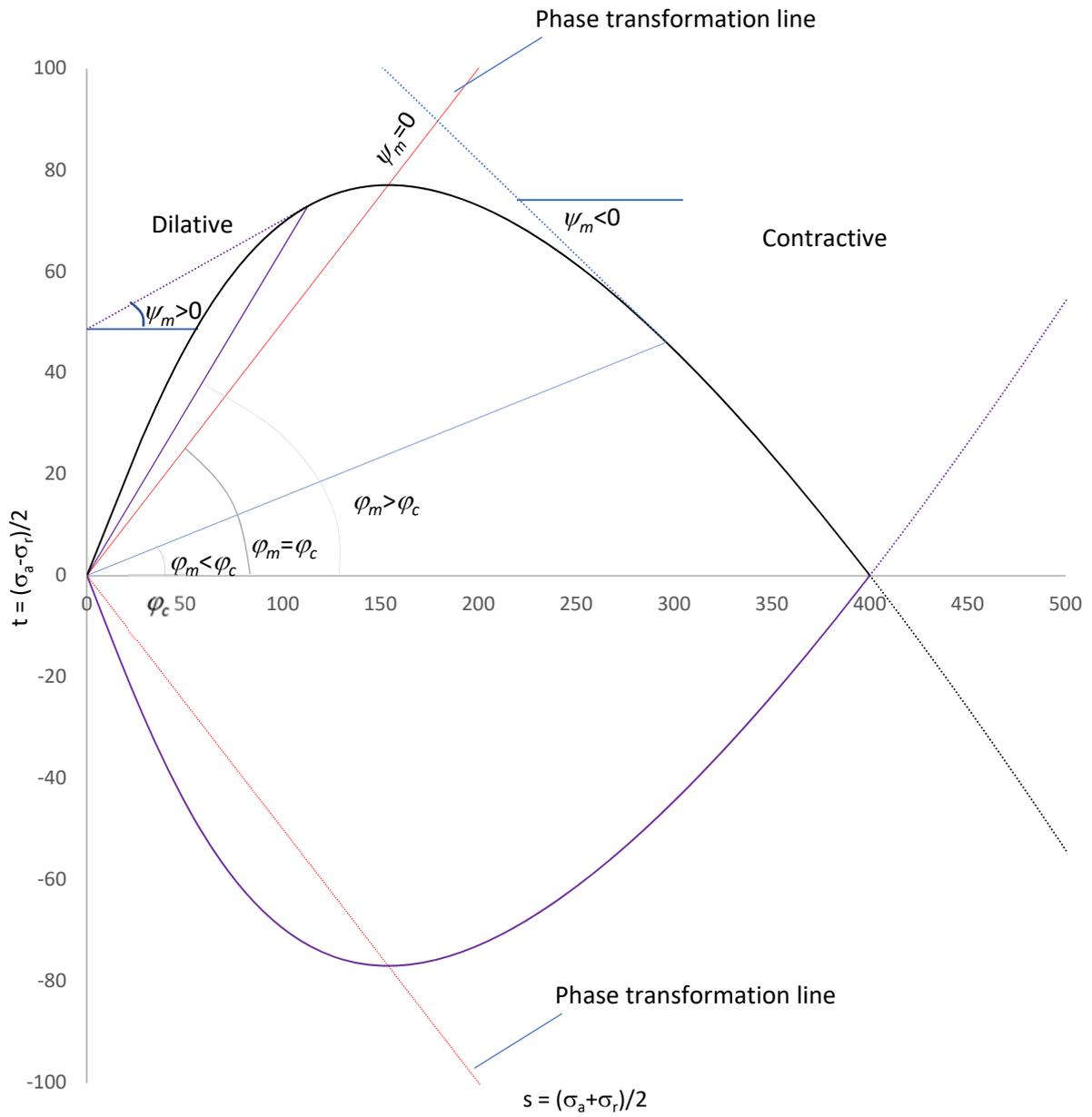

**Figure 5:** Plots ACStD -yield curves in t-s space and $C_N = 3$. Geometric interpretation of dilatancy angle, differentiation of dilative and contractive regions. The solid line is for shear loading and the dotted line is for shear unloading.

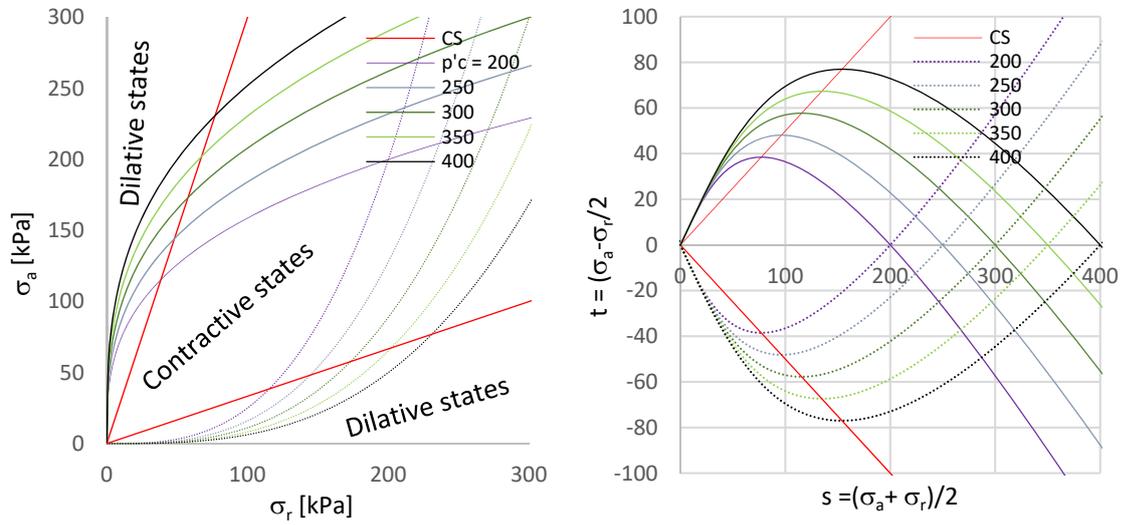

**Figure 6:** ACStD yield curves for loading (solid lines) and unloading (dotted lines) for varying $p_c$ values and constant $C_N = 3$.

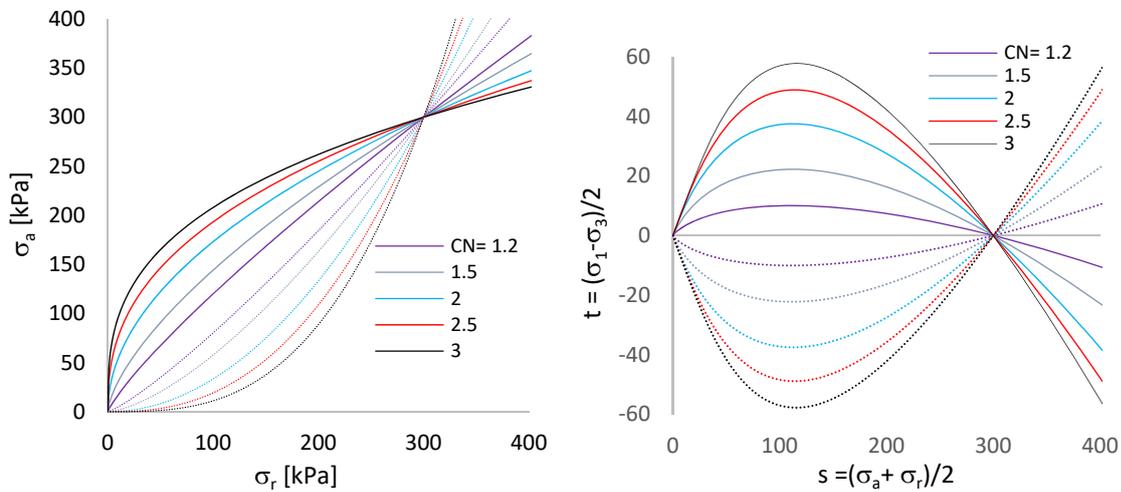

**Figure 7:** ACStD yield curves for loading (solid lines) and unloading (broken lines) for constant $p_c = 300$ kPa and varying $C_N$ values.

## 4.2 Lode angle dependency

So far, we have considered axisymmetric and plane strain conditions. The resulting stress-dilatancy relations do not reflect the effect of intermediate stress state. Here, we will consider other stress invariants which are convenient for establishing stress-dilatancy relations and plastic potential function in the general stress space.

Considering Equation (3) and assuming coaxiality, the plastic dissipation per unit bulk volume can now be written as

$$D^p = (p + a)\dot{\varepsilon}_v^p + s \cos(\theta_\sigma - \theta_{\dot{\varepsilon}}^p) q \dot{\varepsilon}_q^p, \tag{42}$$

where $\theta_\sigma$ is the Lode angle of the effective stress tensor; $\theta_{\dot{\varepsilon}}^p$ is the Lode angle of the plastic strain rate tensor and $s := \mathrm{sgn}(s_{ij}\dot{e}_{ij}^p)$; $s = 1$ for shear loading and $s = -1$ for shear unloading.

As in the plane strain and axisymmetric conditions, one may consider conjugate stress ratio and dilatancy ratio as

$$q = M_\sigma^\theta (p + a) \text{ and } \dot{\varepsilon}_v^p = -M_\psi^\theta \dot{\varepsilon}_q^p, \tag{43}$$

respectively, where $M_\sigma^\theta$ is the stress ratio that depends on the Lode angle (Figure 8) and $M_\psi^\theta$ is the conjugate dilatancy ratio. From Equations (42) and (43), the plastic rate of work per unit volume can be written as

$$D^p = (p + a)\dot{\varepsilon}_q^p \cos(\theta_\sigma - \theta_{\dot{\varepsilon}}^p) d_M, \tag{44}$$

where

$$d_M = -\tilde{M}_\psi^\theta + \tilde{M}_\sigma^\theta, \tag{45}$$

is a function that contains the stress-dilatancy conjugates, $\tilde{M}_\sigma^\theta = s M_\sigma^\theta$ and $\tilde{M}_\psi^\theta = M_\psi^\theta / \cos(\theta_\sigma - \theta_{\dot{\varepsilon}}^p)$

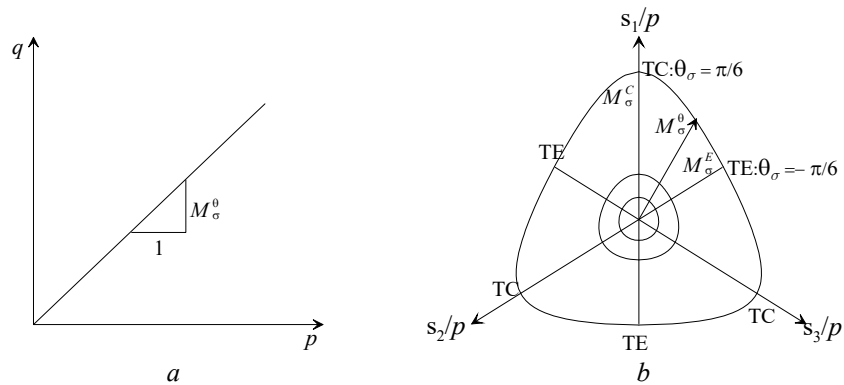

**Figure 8**: Plots of a linear yield line for the intermediate shear mode in *p-q* plane, (b) iso-distortional strain contours in $\pi$-plane, TC: triaxial compression, TE: triaxial extension, $\boldsymbol{\theta_\sigma}$: Lode angle.

Postulating the variation $\delta d_M$ to vanish, *i.e.*, HCSDC [1]:

$$\delta d_M = -\delta \tilde{M}^\theta_\psi + \delta \tilde{M}^\theta_\sigma = 0, \tag{46}$$

yields $d_M$ a constant, say $d_M = C^\theta_M$. This leads to a stress-dilatancy relationship of the form,

$$\tilde{M}^\theta_\psi = \tilde{M}^\theta_\sigma - C^\theta_M, \tag{47}$$

or

$$M^\theta_\psi = \cos(\theta_\sigma - \theta^p_{\dot{\varepsilon}})\left(sM^\theta_\sigma - C^\theta_M\right). \tag{48}$$

$C^\theta_M$ may be evaluated at $\dot{\varepsilon}^p_v \approx \dot{\varepsilon}_v \to 0$, *i.e.*, assuming elastic strain rates to be small. Equation (46) assumes that the sum of the stress ratio and its conjugate dilatancy ratio is a constant, a similar explanation was given by Muir Wood [40].

From Equations (44) and (48), the plastic dissipation is given by

$$\mathcal{D}^p_M = (p+a)C^\theta_M \cos(\theta_\sigma - \theta^p_{\dot{\varepsilon}})\dot{\varepsilon}^p_q \geq 0, \; C^\theta_M \geq 0. \tag{49}$$

When $\theta_\sigma = \theta^p_{\dot{\varepsilon}}$ and $a = 0$, Thurairajah's [41] plastic dissipation equation is obtained. It is also called the Cam clay work hypothesis since it assumes a central place in the plastic potential function of the original Cam clay model. It is also called Taylor's work hypothesis since such type of plastic dissipation was first described by Taylor [11]. Considering the deformation of sand in a direct shear box (See Figure 9 Right), Taylor [11] hypothesized that part of the strain energy used for volume expansion is supplied by a portion of the total shear stress. He further explained that the part of the energy invested in volume expansion is energy spent to overcome interlocking [13]. In simple terms, the total strain energy rate is split into energy expended due to shear and due to dilation. Taylor's hypothesis provided an insight for later thermomechanical developments, for example, [42-47].

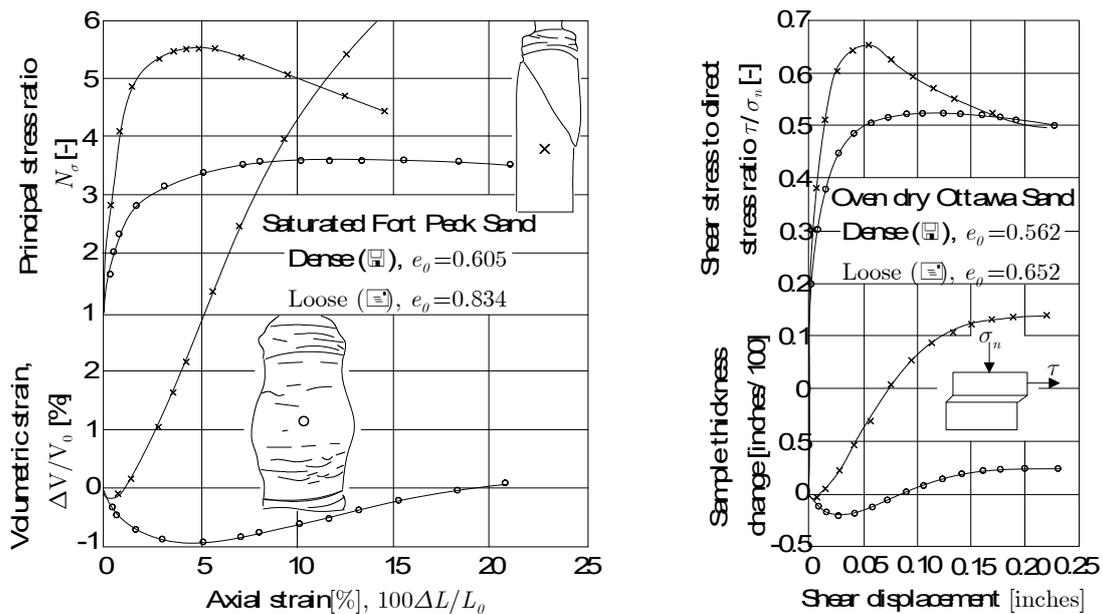

**Figure 9:** Left: Plots of typical triaxial compression tests: Samples of Fort Peck Sand, Right: Typical –plots of direct shear tests: Samples of Ottawa Sand (after Taylor [11])

In Tsegaye [1], it is shown as follows that, for the established framework, plastic unloading in shear is necessarily contractive. Considering the stress-dilatancy relationship in Equation (48), the plastic dissipation can be equivalently written as

$$\mathcal{D}_M^p = (p+a)\dot{\varepsilon}_v^p \frac{C_M^\theta}{C_M^\theta - sM_\sigma^\theta}. \tag{50}$$

For plastic unloading in shear $s = -1$. Therefore, for $C_M^\theta > 0$, $C_M^\theta - sM_\sigma^\theta > 0$ is always valid. This implies that for non-negative plastic dissipation $p\dot{\varepsilon}_v^p \geq 0$ must be satisfied. In soils we deal with compressive stresses (+ for sign convention of soil mechanics), therefore $\dot{\varepsilon}_v^p$ must be positive (*i.e.*, contractive for sign convention of soil mechanics).

In fact, during cyclic undrained simple shear and triaxial compression-extension tests, a significant pore pressure is seen to be generated in the unloading branch. See Figure 10, Figure 11, and Figure 12 for instance.

Let us find the plastic potential function by setting,

$$\frac{M_\psi^\theta}{\cos(\theta_\sigma - \theta_{\dot{\varepsilon}}^p)} = \frac{dq}{dp} = \left(s\frac{q}{p+a} - C_M^\theta\right). \tag{51}$$

This leads to a plastic potential function of

$$q = sC_M^\theta \left[a + p - p\ln\frac{p}{p_{cs}}\right]. \tag{52}$$

where $p_{cs}$ is the effective confining pressure at the critical state, *i.e.*, the critical state is to remain on $q = sC_M^\theta(a+p)$ line. Note that when $s = -1$, the critical state line is defined by an image $q$ which has a negative value. The plastic potential function in Equation (52) is here called Associated Generalized Cyclic Stress Dilatancy yield function, abbreviated AGCStD. Note that when $a = 0$ and $s = 1$ this is just the original Cam clay yield function.

The dilatancy ratio can now be obtained from

$$-\tilde{M}_\psi^\theta = -\frac{dq}{dp} = sC_M^\theta \ln\frac{p}{p_{cs}} \tag{53}$$

Suppose we define the point at which the yield function defined by Equation (52) intercepts the *p*-axis as an apparent pre-consolidation stress, $p_c$, the following relation can be established:

$$\ln p_{cs} = \ln p_c - 1 - \frac{a}{p_c}, \tag{54}$$

or

$$p_{cs} = \frac{p_c}{\exp(1 + \frac{a}{p_c})}, \tag{55}$$

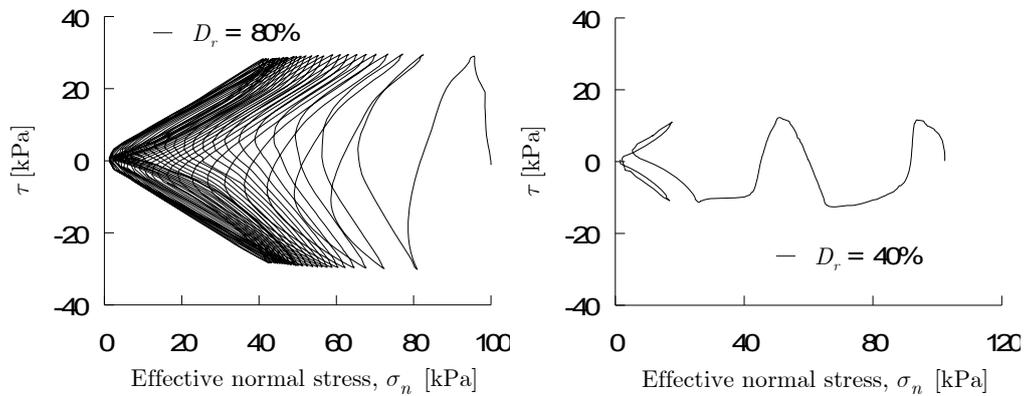

**Figure 10:** Plots of effective normal stress versus shear stress for cyclic undrained simple shear tests on Fraser Delta Sand of varying relative density [48].

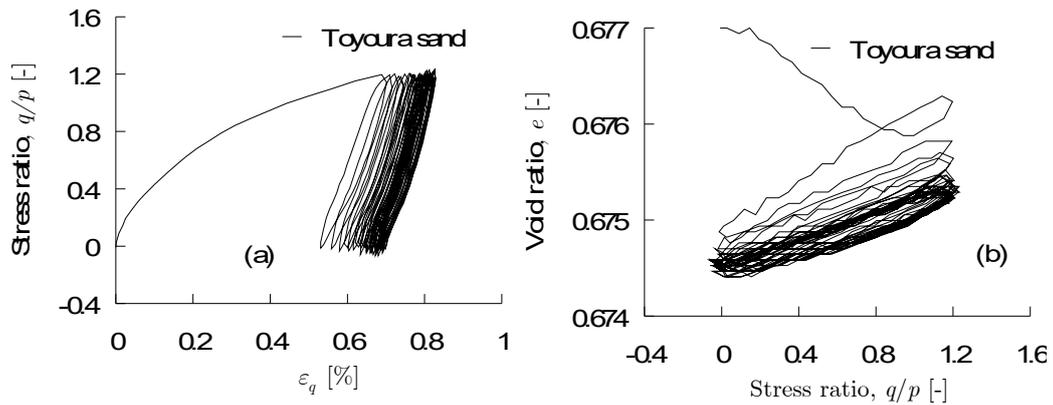

**Figure 11:** Plots of (a) stress ratio versus deviatoric strain and (b) void ratio versus stress ratio at a constant effective confining pressure (at $p$=196 kPa) unloading reloading drained triaxial compression test results on Toyoura Sand [49].

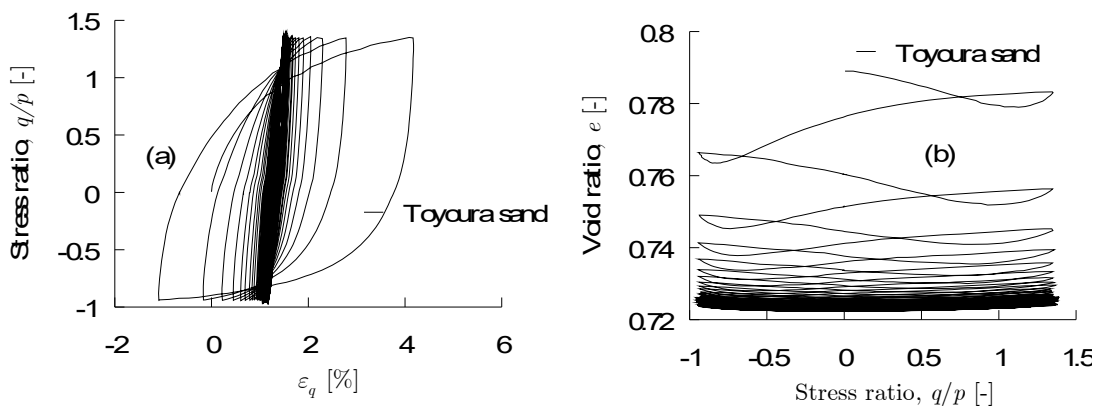

**Figure 12:** Plots of (a) stress ratio versus deviatoric strain and (b) void ratio versus stress ratio of constant effective confining pressure (at $p$=196 kPa) cyclic triaxial compression extension tests under drained conditions of Toyoura Sand [49].

in which $p_c$ is apparent pre-consolidation stress. For a given $p_{cs}$, the corresponding apparent pre-consolidation stress, $p_c$, can also be found as:

$$p_c = \exp\left\{\ln p_{cs} + 1 + W\left(-a\left(\ln p_{cs} + 1\right)\right)\right\}, \tag{56}$$

where $W$ is Lambert's W function.

For $q > 0$ during unloading,

$$q = sC_M^\theta \left(a + p - p \ln \frac{p}{p_{cs}}\right) > 0, \tag{57}$$

or

$$p_{cs} < p \exp\left\{-\frac{a+p}{p}\right\}. \tag{58}$$

When attraction is zero:

$$p_{cs} < p \exp(-1). \tag{59}$$

$C_M^\theta$ may be given as

$$C_M^\theta = 6 f_{sd} \ell_\theta \frac{\sin \varphi_c}{3 - \sin \varphi_c}, \tag{60}$$

where $\varphi_c$ is the critical sate friction angle for triaxial compression condition, $\ell_\theta$ is the Lode angle dependency function and $f_{sd}$ is a density dependency function (given in Equation (40) for instance). For $\ell_\theta = 1$, $C_M^\theta$ defines the stress ratio for the triaxial compression condition according to the Mohr-Coulomb criterion. There are several Lode angle dependency functions that have frequently been applied in constitutive modelling of soils. For instance, Bardet [50] derived a Lode angle dependent function

$$\ell_B^\theta = \frac{\sqrt{3}\omega}{2\sqrt{\omega^2 - \omega + 1}} \frac{1}{\cos \vartheta} \tag{61}$$

for the Matuoka-Nakai criterion, where

$$\omega = \frac{3 - f_{sd} \sin \varphi_c}{3 + f_{sd} \sin \varphi_c} \tag{62}$$

and the angle $\vartheta$ is defined by

$$\vartheta = \frac{\pi}{3}\langle \operatorname{sgn} \theta_{\dot\varepsilon}^p \rangle - \frac{1}{6} \operatorname{sgn} \theta_{\dot\varepsilon}^p \arccos\left(-1 + \frac{27\omega^2(1-\omega)^2}{2(\omega^2 - \omega + 1)^3} \sin^2 3\theta_{\dot\varepsilon}^p\right), \tag{63}$$

in which $\operatorname{sgn} \theta_{\dot\varepsilon}^p = \theta_\sigma / |\theta_{\dot\varepsilon}^p|, \theta_\sigma \neq 0$, is -1 for $\theta_{\dot\varepsilon}^p \leq 0$ and +1 for $\theta_{\dot\varepsilon}^p > 0$ and $\langle . \rangle$ is the Macaulay bracket with a property $\langle a \rangle = a$ if $a > 0$ and $\langle a \rangle = 0$ if $a \leq 0$. Note that, here the Lode angle is to be defined

by the direction of the plastic strain increment instead of the stress. For the case of proportional monotonic loading, the Lode angle of the effective stress and that of the Lode angle of the plastic strain increment are equal.

The solutions provided here are generalization of the original Cam clay yield function. One of the limitations constitutive modellers looked away from the original Cam clay yield function is the direction of plastic strain increments during isotropic compression-which is considered unphysical. A new insight is obtained with the consideration of unloading. That is, it can be clearly seen that the loading and unloading AGCStD curves are intersecting along the isotropic axis and for any plastic compression along this axis both are equally likely. In other words, one direction is not any preferable than the other and uniqueness of the direction of plastic flow direction is lost for the stress state at this point. We may consider then Koiter's rule [51] and sum the plastic strain increments in both direction which gives us

$$\left(-\frac{dq}{dp}\right)_L + \left(-\frac{dq}{dp}\right)_U = C_M^\theta \ln\frac{p_c}{p_{cs}} - C_M^\theta \ln\frac{p_c}{p_{cs}} = 0. \tag{64}$$

Example plots of AGCStD yield curves for loading and unloading -and constant critical state friction angle and varying apparent pre-consolidation stress. The extension of the curves beyond the pre-consolidation stress is valid only for unloading.

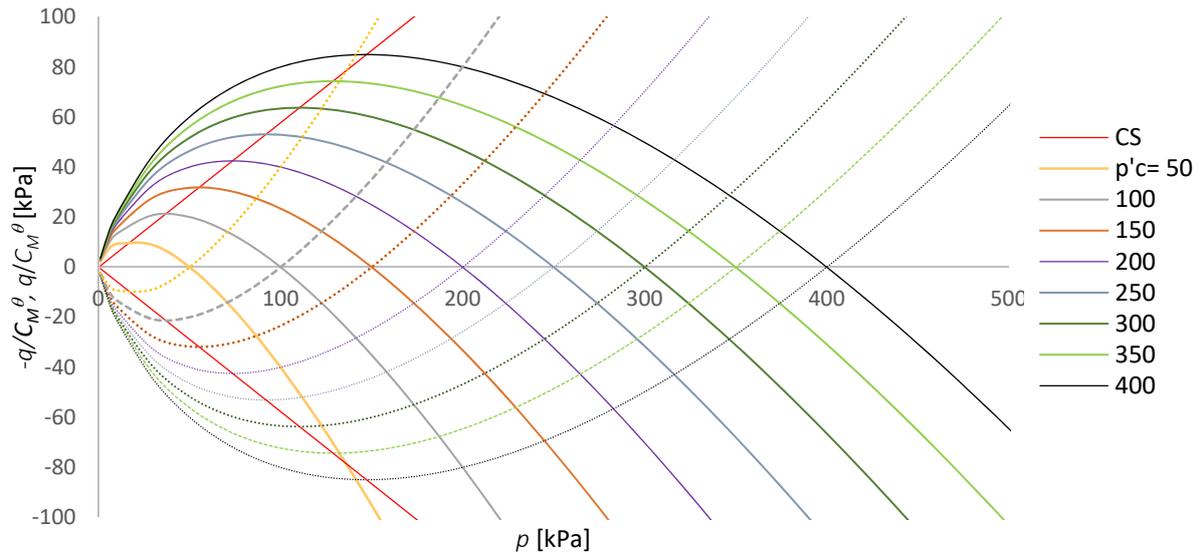

**Figure 13:** Plastic potential functions described by Equation (52). Solid lines are for loading and dotted lines are for unloading. Both the unloading and the loading are continued to an imaginary space defined by $-q/C_M^\theta$.

The theory can also be applied directly considering the shear stress and the normal stress along the Matsuoka-Nakai spatial mobilized plane, Figure 14. Such a consideration leads us to a yield function, of the form:

$$f = \sqrt{I_{1\sigma}I_{2\sigma}I_{3\sigma} - 9I_{3\sigma}^2} - \frac{2}{3}sf_{sd}\tan\varphi_c\left\{a - 3I_{3\sigma}\left(\ln\left(\frac{3I_{3\sigma}}{I_{2\sigma}p_c}\right) + \frac{a}{p_c}\right)\right\} = 0, \tag{65}$$

which is here called AGCStD-MN where $I_{1\sigma}$, $I_{2\sigma}$ and $I_{3\sigma}$ are the first, the second and the third stress invariants which are given in terms of the principal stresses respectively as $I_{1\sigma} = \sigma_1 + \sigma_2 + \sigma_3$, $I_{2\sigma} = \sigma_1\sigma_2 + \sigma_2\sigma_3 + \sigma_3\sigma_1$ and $I_{3\sigma} = \sigma_1\sigma_2\sigma_3$.

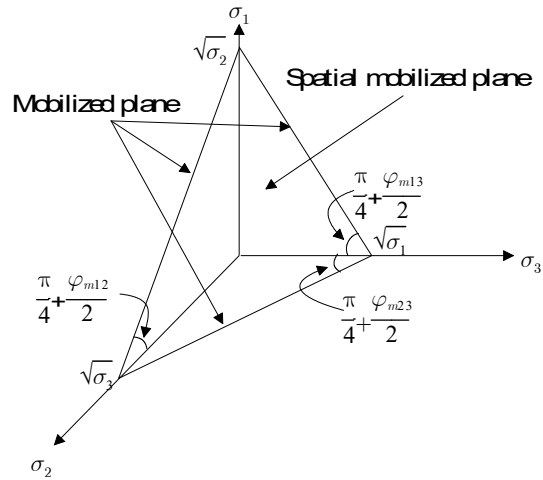

**Figure 14:** The Matusoka-Nakai spatial mobilized plane [52]

# 5. Summary and conclusions

In this paper, yield functions are derived for frictional materials based on stress-dilatancy relationships within the framework of associated plasticity. The treatise begins by presenting the energy variation due to deformation of a continuum body for an isothermal condition. Considering the additive decomposition of strain rates into elastic and plastic, the rate of work of a deforming body is further split into an elastic (stored) and a plastic (dissipated) component. The plastic dissipation is rearranged in such a way stress-dilatancy conjugates are identified. Conjugate stress ratio and dilatancy ratio are related to each other through the hypothesis of *complementarity of stress-dilatancy conjugates* (HCSDC) and the results are further used for establishing yield functions in the framework of associated plasticity. Considering non-negative rate of plastic dissipation, the theoretical framework is setup for both loading and unloading - where loading away from isotropy is defined as loading in shear and loading towards isotropy is defined as unloading in shear. With the consideration of both loading and unloading in shear, the theory aims at a generalized framework for the modelling of the deformation behavior of soils subjected to not only of monotonic loading but also cyclic loading. This consideration is also shown to give a new insight into the sharp front for the original Cam clay model which constitutive modelers consider as an undesired property. The theoretical framework may prove useful for application in constitutive modelling of soils subjected to monotonic and cyclic loading and reconsidering some of the axioms of associated plasticity in relation to the mechanical behavior of soils. The current treatment is limited to the assumption of coaxiality between eigen directions of stresses and plastic strain rates. The author wishes to follow up the current exposition with applications and extension of the theory to non-coaxial plastic flow and anisotropic conditions.